\documentclass[journal,twoside,web]{ieeecolor}
\usepackage{tuffc}
\usepackage{cite}
\usepackage{amsmath,amssymb,amsfonts}
\usepackage{algorithmic}
\usepackage{graphicx}
\usepackage{textcomp}

\usepackage{soul,color}

\usepackage{multirow}
\usepackage{booktabs}
\usepackage{wrapfig,colortbl}
\usepackage[pagewise,switch]{lineno}
\usepackage[hidelinks]{hyperref}
\soulregister\cite7
\soulregister\ref7
\soulregister\pageref7
\soulregister\eqref7

\definecolor{abstractbg}{rgb}{1,0.969,0.914}
\def\BibTeX{{\rm B\kern-.05em{\sc i\kern-.025em b}\kern-.08em
    T\kern-.1667em\lower.7ex\hbox{E}\kern-.125emX}}
\begin{document}
\title{Vector Flow Imaging in Layered Models With a High Speed of Sound Contrast Using Pulse-Echo Ultrasound and Photoacoustics}
\author{Caitlin Smith, Guillaume Renaud, Kasper van Wijk, and Jami Shepherd
\thanks{This work is funded by the Royal Society of New Zealand Marsden Fund project number MFP-UOA2014.}
\thanks{C. Smith, K. van Wijk and J. Shepherd are with the Department of Physics, The University of Auckland, Auckland 1010, New Zealand and also with the Dodd-Walls Centre for Photonic and Quantum Technologies, Auckland, New Zealand (e-mail: c.g.smith@auckland.ac.nz, k.vanwijk@auckland.ac.nz and jami.shepherd@auckland.ac.nz). J. Shepherd is also with the MacDiarmid Institute for Advanced Materials and Nanotechnology, Department of Physics, University of Auckland, Auckland 1010, New Zealand.}
\thanks{G.R.R Renaud is with the Department of Imaging Physics, Delft University of Technology, Delft, 2628 CN, The Netherlands  (e-mail: G.G.J.Renaud@tudelft.nl).}}

\IEEEtitleabstractindextext{%
\fcolorbox{abstractbg}{abstractbg}{%
\begin{minipage}{\textwidth}\rightskip2em\leftskip\rightskip\bigskip
\begin{wrapfigure}[15]{r}{4in}
\hspace{-3pc}\includegraphics[width=4in]{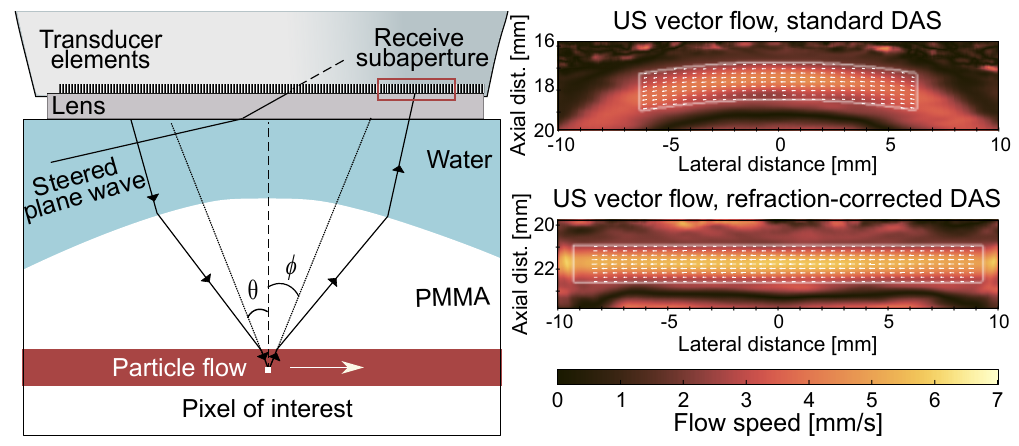}
\end{wrapfigure}%
\begin{abstract}

 In this study, we develop vector flow imaging techniques for multi-layered models with a high wavespeed contrast using photoacoustic and ultrasonic imaging. We use refraction-corrected delay-and-sum image reconstruction (RC-DAS), which enforces Snell’s law to accurately calculate time delays within each layer. We compare RC-DAS against conventional delay-and-sum for vector flow imaging in benchtop phantoms made of transparent polymethyl methacrylate (PMMA) in a water bath. We study the flow beneath a PMMA layer using two phantoms, where the PMMA layer has different shapes and thicknesses. We image a slow-moving suspension of carbon microspheres ($\sim$4~mm/s) using interleaved photoacoustic and multi-angle plane wave ultrasound acquisitions measured with a 7.6~MHz linear ultrasound array. Photoacoustic waves are generated by a 1064~nm wavelength nanosecond-pulsed laser at 50~Hz, and multi-angle plane wave ultrasound data are acquired at 100~Hz for eleven steering angles between $\pm$10$^\circ$. RC-DAS improves the flow speed accuracy, reducing the mean absolute error by 0.41-0.63~mm/s compared to the expected flow profile. The error in direction estimates improves when we use RC-DAS, with the interdecile range reducing by up to 17$^\circ$. This work emphasises the importance of refraction correction for accurate flow measurements in layered media with photoacoustics and ultrasonic imaging. While both imaging modalities can quantify flow in these multi-layered models, the modality best suited for a specific application will depend on the imaging target and flow dynamics. These techniques show promise for biomedical applications such as intraosseous and transcranial blood flow quantification, and in nondestructive testing to monitor fluid motion.
\end{abstract} 


\bigskip
\end{minipage}}}

\maketitle

\section{Introduction}
\IEEEPARstart{U}{ltrasound} (US) imaging of flowing particles, such as blood, is well-established and routinely used clinically. A high-frequency (MHz) acoustic wave is transmitted by an US probe and scattered by structures in a medium. These scattered waves are detected by the US probe, enabling images showing the location of the scatterers to be generated. In contrast, in photoacoustic (PA) imaging, a pulse of laser light is absorbed by optical absorbers within the medium, resulting in transient heating and a subsequent pressure wave that propagates from the optical absorbers. In acoustic-resolution PA imaging, these acoustic waves are detected by an US probe, allowing images showing the location of the optical absorbers to be obtained. Compared to US imaging, in which both the transmit and receive components travel as an acoustic wave, PA imaging only has one-way acoustic propagation.

\begin{table*}[!t]
\arrayrulecolor{subsectioncolor}
\setlength{\arrayrulewidth}{1pt}
{\sffamily\bfseries\begin{tabular}{lp{6.75in}}\hline
\rowcolor{abstractbg}\multicolumn{2}{l}{\color{subsectioncolor}{\itshape
Highlights}{\Huge\strut}}\\
\rowcolor{abstractbg}$\bullet$ & We present a method for vector flow imaging of flowing particles in benchtop multi-layered models with a strong speed of sound contrast for pulse-echo ultrasound and photoacoustic imaging.\\
\rowcolor{abstractbg}$\bullet${\large\strut} & Compared to standard delay-and-sum beamforming, refraction-corrected delay-and-sum significantly improves the accuracy of the image geometry and vector flow images.\\
\rowcolor{abstractbg}$\bullet${\large\strut} & Correcting for the acoustic refraction at the layer interfaces is essential for accurate vector flow imaging with ultrasound and photoacoustics.
 \\[2em]\hline
\end{tabular}}
\setlength{\arrayrulewidth}{0.4pt}
\arrayrulecolor{black}
\end{table*}

Power Doppler is commonly used in clinical US imaging to visualise blood flow and quantify blood volume by estimating the image power after clutter filtering to isolate the flowing component~\cite{rubin}. However, Power Doppler does not provide quantitative flow information such as flow speed or direction. Colour Doppler utilises the Doppler shift to quantify flow in the direction of the US beam. To determine the absolute flow speed, the clinician must choose an appropriate transmit beam angle and determine the vessel angle, which requires the geometry to be simple and clearly visible.

Vector flow imaging (VFI) advances on colour Doppler by detecting the Doppler shift from multiple beam angles to quantify flow, making VFI less reliant on user-defined parameters and more suitable for imaging tortuous flow geometries. VFI techniques to quantify the speed and direction of flowing particles within a single speed of sound (SOS) medium are well-established in biomedical ultrasound~\cite{yiu_least-squares_2016, goddi_vector_2017, jensen_ultrasound_2016-1, jensen_ultrasound_2016} and emerging in photoacoustic imaging~\cite{van_den_berg_review_2015, pakdaman_zangabad_photoacoustic_2021, zhang_photoacoustic_2024, smith_vector-flow_2024}. These state-of-the-art VFI techniques analyse reconstructed images over time to quantify motion. However, the conventional image reconstruction approach assumes a single SOS in the medium. This approach is inaccurate in situations with high SOS contrast, such as in layered media, where the different SOS in each layer causes refraction of acoustic waves at the interfaces between layers~\cite{renaud_vivo_2018, shepherd_photoacoustic_2020}.

In this work, we present a method for quantifying particle flow at every pixel with both PA vector flow (PAVF) and US vector flow (USVF) in multi-layered models with a high SOS contrast. We correct for refraction in the reconstruction of the US \cite{renaud_vivo_2018} and PA images~\cite{shepherd_photoacoustic_2020}, before applying a multi-angle vector flow algorithm to estimate the velocity at every pixel~\cite{smith_vector-flow_2024}. These techniques are validated on PA and pulse-echo US data from multi-layered phantoms containing a flowing carbon suspension to simulate flow within and beneath the high SOS layer made from polymethyl methacrylate (PMMA). Using refraction-corrected image reconstruction, the accuracy of the B-mode images and the vector flow images with PA and US are reported, and compared to standard delay-and-sum reconstruction (DAS). These techniques are promising for biomedical applications, such as quantifying intraosseous flow, and transcranial applications, as well as in nondestructive testing (NDT) to monitor fluid motion.

\section{Principles of refraction-corrected vector flow imaging}
\subsection{Homogeneous speed of sound delay-and-sum imaging}

DAS image reconstruction, commonly used in both PA and US imaging, calculates travel times ($t$) based on the SOS in a medium. This travel time includes two components: the transmission delay (source to pixel) and the receive delay (pixel to receiver). For PA, only the latter is considered. The pressures detected by each transducer element are mapped to the spatial imaging grid based on these time delays, and the images obtained from each element are summed together. Conventional DAS assumes a uniform SOS, so the US delay is the source-pixel-receiver distance divided by the SOS. For PA, the delay is simply the pixel to receiver distance, divided by the SOS.

\subsection{Refraction-corrected delay-and-sum imaging}\label{RC_section}

\begin{figure}
    \centering
    \includegraphics{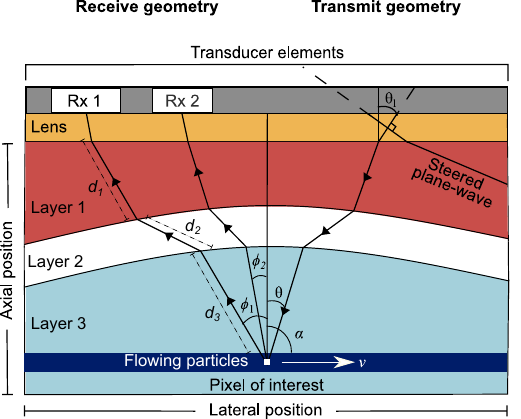}
    \caption{Refraction of acoustic rays to and from an US scatterer (indicated as the ``pixel of interest''). For the plane wave transmission, the steering angle in the lens ($\theta_l$) is chosen by the operator, while the transmit angle at the pixel $\theta$ is calculated and used in \eqref{matrix_vf_US}. The receive angle at a pixel $\phi$ is changed using multiple receive subapertures (labelled ``Rx'') to obtain the desired $\phi$ angles during reconstruction.}
    \label{geo}
\end{figure}

However, in layered media with a high SOS contrast, there is significant refraction that occurs at the interfaces between layers (Fig.~\ref{geo}). In this case, assuming a homogeneous SOS results in inaccurate localisation within the image and aberration artefacts. For example, in clinical US imaging of soft tissues, such as muscle (1580~m/s) and fat (1450~m/s), the SOS varies less than 10\% from the assumed average SOS of 1540 m/s~\cite{szabo_appendix_2014}, so assuming straight ray paths is reasonable. However, since the SOS in cortical bone ($\sim$3500~m/s \cite{granke_change_2011, renaud_vivo_2018}) is significantly higher than the soft-tissue SOS, standard DAS is unsuitable for imaging within or beneath bone layers as the large SOS mismatch between soft-tissue and bone causes refraction at the interfaces. A summary of refraction correction techniques used in clinical US imaging is described in the review by Ali \textit{et al.}~\cite{ali_aberration_2023}, and the work by Mozaffarzadeh \textit{et al.} compares two common approaches~\cite{mozaffarzadeh_comparison_2022}.

Our method for refraction-corrected DAS (RC-DAS) is based on Kirchhoff migration from seismology~\cite{waltham_two-point_1988,etgen_overview_2009, renaud_real-time_2018}, which corrects the propagation path and time delays to accurately reconstruct images in layered models. RC-DAS requires the location of the interfaces and the SOS in each layer to be known or estimated a priori to image reconstruction~\cite{renaud_vivo_2018}. In our approach, we automatically identify and segment the layer interfaces using US. The receive delay is found by launching rays from the transducer element at a range of angles and finding the fastest path to the pixel. These rays refract at the interfaces based on Snell's law, and the delay ($t$) for each ray is found as 

\begin{equation} 
   t=\sum_{i=1}^N\frac{d_i}{c_i}, 
\end{equation} 

where $i$ indexes each layer of the medium, and $d_i$ and $c_i$ are the distance and SOS in these layers, respectively. Similarly, for US, the transmit delay is calculated by launching rays from a virtual source used to generate a plane wave at a range of different angles and determining the fastest ray path from the source to the pixel (Fermat's principle). To improve the accuracy of these delay calculations, we also consider refraction between the transducer lens and the medium when computing these delays~\cite{waasdorp_assessing_2024}. 

\subsection{Vector flow algorithm}
To compute the magnitude and direction of flow at every pixel, we employ a vector flow approach based on multi-angle observations \cite{smith_vector-flow_2024, yiu_least-squares_2016}. This technique requires a series of US and PA images over time that are each reconstructed from multiple receive angles generated by using smaller subsets of transducer elements for image reconstruction. For US, different plane wave steering angles are also used. The receive and transmit schemes are referred to as ``Rx'' and ``Tx'', respectively. 

\subsubsection{Photoacoustic vector flow}

As described in our previous work \cite{smith_vector-flow_2024}, movement of an optical absorber through a location corresponding to a pixel at ($x,z$) results in a relative phase shift ($\Delta \Psi /2\pi$) between two consecutive frames. This phase shift is related to the displacement of the absorber between frames, $U$, via the SOS in the medium $c$, and the detected centre frequency of the PA waves $f_{0}$. This is described by the so-called PA Doppler equation:

\begin{equation}\label{pa_dopp_1}
     \frac{\Delta \Psi}{2\pi}=\frac{Uf_{0}}{c}\cos(\phi-\alpha),
\end{equation}
where $\phi$ is the angle of the receive beam at the pixel and $\alpha$ is the flow angle, both measured relative to the transducer surface normal (see Fig.~\ref{geo}). Using the difference of cosines, we can express this equation in terms of the axial and lateral flow components ($U_z=U\cos \alpha$, $U_x=U\sin \alpha$):

\begin{equation}\label{pa_dopp_2}
     \frac{\Delta \Psi}{2\pi}=\frac{f_{0}}{c}\left[U_z\cos\phi+U_x\sin\phi\right]. 
\end{equation}

The average phase shift over an ensemble of consecutive frames $\overline{\Delta \Psi}$ is estimated at each pixel using the lag-one autocorrelation common to ultrasonic colour flow imaging ~\cite{namekawa_realtime_1983, kasai1985real, evans_ultrasonic_2011, evans_colour_2010, jensen_estimation_1996}. By estimating $\overline{\Delta \Psi}$ for $n$ different Rx, we obtain estimates of the phase shift for different $\phi$. By using three or more Rx, an over-determined system of linear equations can be generated:
 
\begin{equation}\label{matrix_vf_PA}
\begin{bmatrix}
\cos \phi_1 & \sin \phi_1\\
\vdots &\vdots\\
\cos \phi_n & \sin \phi_n\\
\end{bmatrix}
\begin{bmatrix}
U_z\\
U_x 
\end{bmatrix}
=
\frac{c}{2\pi f_{0}}
\begin{bmatrix}
\overline{\Delta \Psi}_1\\
\vdots\\
\overline{\Delta \Psi}_n
\end{bmatrix}.
\end{equation}

Since \eqref{matrix_vf_PA} is a system of linear equations of the form $A\mathbf{x}=\mathbf{b}$, we can estimate the displacement components contained in $\mathbf{x}=[U_z U_x]^T$ by minimising $\|A\mathbf{x}-\mathbf{b}\|$ using a least-squares approach. For a known frame rate, FR, the displacement per frame $U$ can be converted into a velocity $v=U\times$ FR. For PA imaging, the FR is given by the laser pulse repetition frequency. 

\subsubsection{Ultrasound vector flow}
 Analogous to the PA example, a scatterer moving at a velocity $v$ and angle $\alpha$ through a location corresponding to a pixel $(x, z)$ will generate a local relative phase shift ($\Delta \Psi /2\pi$) between two consecutive frames. $\Delta \Psi$ is related to the displacement per frame $U$ via the US Doppler equation: 

\begin{equation}\label{US_dopp_1}
     \frac{\Delta \Psi}{2\pi}=\frac{Uf_0}{c}\left[\cos(\theta-\alpha)+\cos(\phi-\alpha)\right], 
\end{equation}

where $\theta$ is the angle of the transmit beam at the pixel. Using the difference of cosines trigonometric identity and substituting in the axial and lateral displacement components, we obtain the following:

\begin{equation}\label{US_dopp_2}
     \frac{\Delta \Psi}{2\pi}=\frac{f_0}{c}\left[U_z\left(\cos\theta+\cos\phi\right)+U_x\left(\sin\theta+\sin\phi\right)\right]. 
\end{equation}

Like the PA analogue, using $n$ receive angles and $m$ steered plane wave transmissions, we can estimate $\overline{\Delta\Psi}$ for the $n\times m$ Tx/Rx combinations. From this, we obtain an over-determined linear equation shown in Equation~\eqref{matrix_vf_US}, which can be solved using a least-squares approach introduced by Yiu \textit{et al.} for USVF~\cite{yiu_least-squares_2016}.

\begin{equation}\label{matrix_vf_US}
\begin{bmatrix}
\!\cos \theta_1 + \cos \phi_1 \!\! & \sin \theta_1+\sin \phi_1\!\\
\vdots &\vdots\\
\!\cos \theta_m + \cos \phi_n \!\!& \sin \theta_m+\sin \phi_n\!\\
\end{bmatrix}\! \!
\begin{bmatrix}
U_z\\
U_x 
\end{bmatrix}\!\!
=\!
\frac{c}{2\pi f_0}\!
\begin{bmatrix}
\overline{\Delta \Psi}_1\\
\vdots\\
\overline{\Delta \Psi}_{mn}
\end{bmatrix}
\end{equation}

\section{Experimental Methods}
\begin{figure}
    \centering
    \includegraphics{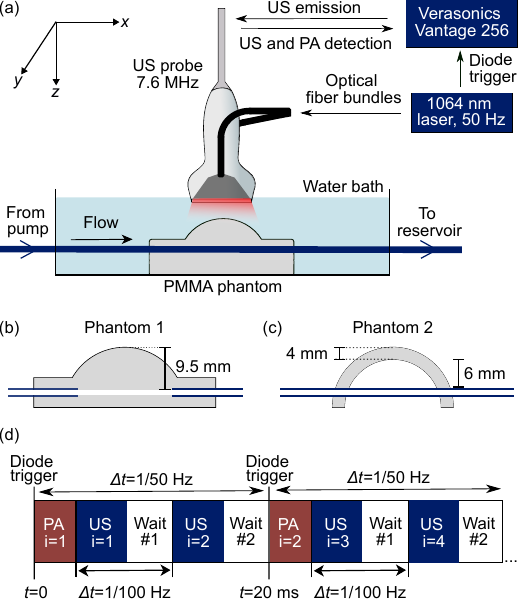}
    \caption{(a) Experimental setup for interleaved PA and US acquisitions of the multi-layered models. The carbon particle suspension is pumped through the channel via a syringe pump. The US transducer is suspended at the surface of the water bath, and the laser light is directed to the phantoms via an optical fibre bundle, which has two linear outputs on either side of the US probe. Cross-sections of the PMMA phantoms studied in these experiments are shown in (b-c). The blue lines show the walls of the optically-transparent PVC tubing, which transfers the carbon suspension from the syringe pump through the channel. (d) Timing sequence for interleaved acquisition of PA and US frames. Each frame recorded by each modality is indexed by “i”. Red boxes correspond to the acquisition of a PA frame from a single laser pulse, while each blue box contains one US frame composed of eleven steered plane wave transmissions. The width of the boxes is not to scale. ``wait \#1'' ensures the US FR is 100~Hz, while ``wait \#2'' is the time until the next PA acquisition so that the PA FR is 50~Hz.}
    \label{setup}
\end{figure}

\subsection{Experimental setup}

\begin{figure*}
    \centering
    \includegraphics{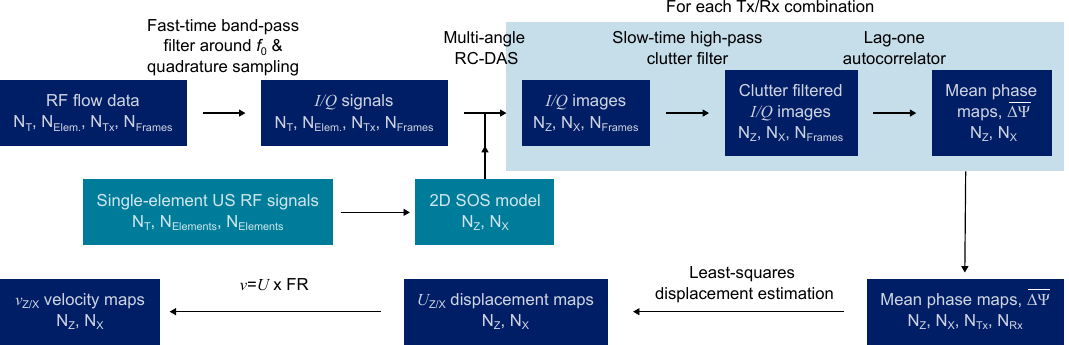}
    \caption{Data processing pipeline for USVF imaging using RC-DAS. The dimensions are: N$_\text{T}$ - the number of fast time samples, N$_\text{Elem.}$ - the number of transducer elements, N$_\text{x/z}$ - the size of the images in the $x/z$ direction, N$_\text{Tx}$ - the number of steered plane waves, and N$_\text{Rx}$ - the number of receive angles. For PAVF, the dimension N$_\text{Tx}$ is removed.}
    \label{process}
\end{figure*}

We designed two phantoms to simulate flow within and beneath the high-contrast SOS layer made from PMMA, referred to as Phantom 1 and Phantom 2, respectively (Fig~\ref{setup}b-c). These phantoms could be considered to mimic blood flow in cortical bone and in bone marrow, respectively, or flow inside a pipe for NDT applications. Both phantoms are milled using computer numerical control machining from a 2.5-cm-thick sheet of optically-transparent PMMA with a SOS of 2750~m/s, which was estimated by measuring the two-way travel time through a 1-cm-thick piece of PMMA with US. This SOS is 46\% faster than the SOS surrounding water bath $c=1485$~m/s, which was found via PA imaging of a point target using an auto-focussing approach~\cite{treeby_automatic_2011}. For reference, the SOS in cortical bone is typically between 2800-4200~m/s \cite{granke_change_2011, renaud_vivo_2018}. The acoustic attenuation coefficient of PMMA is $\sim$1~dB/cm/MHz~\cite{ono_comprehensive_2020}. The phantoms feature a curved outer surface (radius of curvature = 20~mm) to induce strong aberration. Both phantoms are connected to the syringe pump via optically-transparent PVC polymer tubing (inner diameter 1/16" inches, outer diameter 1/8 inches, Sigma Aldrich), shown as the blue lines in Fig~\ref{setup}b-c.

We pump a suspension of carbon particles (1\% v/v, 2-12 $\mu$m diameter, Sigma Aldrich) in a sodium polytungstate solution ($\rho=$~1.5~g/cm$^3$) through the phantoms at a rate of 7.9~$\pm~0.1~\mu$L/s ~\cite{fang_photoacoustic_2007}. Since the channel of Phantom 1 was drilled by hand, the diameter has a lower precision and is slightly larger than the inner diameter of the tubing used in the Phantom 2 ($1.65~\pm~0.05$~mm for the Phantom 1, compared to 1/16" or $1.59~\pm~0.01$~mm for Phantom 2). This results in an average flow speed of $3.70~\pm~0.28$~mm/s in Phantom 1 and $3.99~\pm~0.07$~mm/s in Phantom 2.  

\subsection{Data acquisition system}
For transmission of US plane waves and detection of PA and US signals we use the L11-5v transducer which has a centre frequency of $f_0=$~7.6~MHz, controlled by the Verasonics Vantage 256 system (Verasonics, Kirkland, WA). For PA generation, we use a diode-pumped solid state laser at a wavelength of 1064~nm  (Centurion$+$, Quantel by Lumibird, France) with a 13~ns pulse width (FWHM) and a repetition frequency of 50~Hz. The light is coupled into an optical fiber bundle with two linear outputs positioned on each side of the US probe to illuminate the phantoms with a combined pulse energy of $\sim 25~\text{mJ}/\text{cm}^2$. 

We generate steered plane waves using a virtual source approach to calculate the transmit delay for each element. Eleven steered plane waves with voltage $V_p=20$~V and a pulse length of two half-cycles are transmitted for each US frame, separated by 500~$\mu$s between the different plane wave angles. The steering angles of the plane waves (measured in the lens, $\theta_l$, see Section~\ref{multi_angle}) spanned 20$^\circ$~\cite{yiu_least-squares_2016}. 

We developed a specialised imaging script to enable interleaved acquisitions of PA and US frames, illustrated in Fig.~\ref{setup}d. In this sequence, we acquire two US frames for each PA frame, resulting in an US FR of 100~Hz. This approach ensures that PAVF and USVF estimate the same maximum flow speed without experiencing aliasing, assuming the most extreme $\theta$ is the same as the most extreme $\phi$, and the centre frequencies for PA and US are the same. The aliasing limit occurs when $\Delta\Psi/2\pi=\pm0.5$, so by substituting this into \eqref{pa_dopp_1} and \eqref{US_dopp_1} and using the largest $\phi$ and $\theta$ angles which will experience aliasing at the lowest flow speeds, the maximum measurable speed is 4~cm/s for PAVF ($\phi_{max}=10^\circ$) and 2~cm/s for USVF ($\phi_{max}=10^\circ$ and $\theta_{max}=19^\circ$ found inside the channel of Phantom 1. Finally, we estimate the 2D three-layer model needed for RC-DAS with a full-matrix capture of US data immediately before the simultaneous PA/US acquisition sequence. 

\begin{table}
\caption{Parameters used for PAVF and USVF}
\begin{tabular}{@{}lll@{}}
\toprule
                                                    & PAVF & USVF \\ \midrule
        Frame rate, Hz & 50 & 100\\
        Number of frames & 500 & 1000\\ 
        Clutter filter cut-off, Hz & 2.5 & 5\\
        Transmit angles (in the lens, $\theta_l$) & N.A. & $\pm~10^\circ$, 11 Tx \\
        Receive angles (at the pixel, $\phi$) & $\pm~10^\circ$, 5 Rx & $\pm~10^\circ$, 5 Rx\\
        Centre frequency $f_0$, MHz & 5.37 & 7.60 \\ 
        Fast-time band-pass filter, MHz& 3.37-7.37 & 6.08-9.12 \\ 
        Estimated FWHM in $y$-direction, mm & 1.1 & 0.8\\
        Axial resolution, mm & 0.33 & 0.33\\
        Lateral resolution, mm & 1.33 & 0.94\\
        Vector flow axial resolution, mm & 0.36 & 0.36 \\ \hline
        \multicolumn{3}{c}{}\\
    \multicolumn{3}{p{240pt}}{The transmit angles shown in the table above are measured in the lens ($\theta_l$). The US cut-off frequencies for the fast-time band-pass filter correspond to [0.8 1.2]$\times f_0$. The explanation for finding the PA centre frequency can be found in Section~\ref{PA_fo_disc}.}
    \end{tabular}
    \label{table_params}
\end{table} 

\newpage
\section{Data processing}
\color{black}

In the following section, we detail the processing
workflow beginning with the RF PA and US data through to generating quantitative vector flow images in the multi-layered models. This is summarised by Fig.~\ref{process}.

\subsection{Generating the three-layer model}
To perform RC-DAS, the SOS in each layer and the location of the interfaces must be known. In this study, the SOS for each layer is already known, however, these quantities can be found using an autofocusing approach~\cite{treeby_automatic_2011, renaud_measuring_2020} or by using the headwave velocity to estimate the SOS in the PMMA layer~\cite{bossy_effect_2002}. 

We follow the method presented by Renaud~\textit{et al.} \cite{renaud_vivo_2018} for isotropic media to obtain the three-layer model using the full-matrix capture data. In Phantom 2, we consider the water below the PMMA layer and the tubing to be one layer and assume the SOS of the carbon suspension is the same as the water bath SOS (1485~m/s). We locate the PMMA layer interfaces using the process below: 

\begin{enumerate}
    \item Standard DAS image reconstruction of the entire image using $c_{\text{water}}=$~1485~m/s.
    \item Use the Dijkstra path-finding algorithm to automatically identify the outer PMMA interface between the water and PMMA layer~\cite{hong_medical_2012}. Fit the segmented signal to a second-degree polynomial.
    \item Use RC-DAS to reconstruct the pixels beneath the outer PMMA interface using $c_{\text{PMMA}}=$~2750~m/s.
    \item Repeat step (2)-(3) to identify the inner interface between the PMMA and water/channel layers, then reconstruct the water/channel layer beneath $c_{\text{water}}=$~1485~m/s.
\end{enumerate}

\subsection{Preprocessing}\label{PA_fo_disc}
We band-pass filter the RF PA and US signals around their respective centre frequencies before quadrature sampling the signals to obtain the real ($I$) and imaginary ($Q$) components via the Hilbert transform. The centre frequencies and the filter cut-off frequencies are stated in Table~\ref{table_params}. Determining the PA centre frequency is not straight-forward since the emitted PA spectrum is broadband and convolved with the transducer frequency response; therefore, there can be a mismatch between the detected PA spectrum and the transducer centre frequency. Additionally, the PA spectrum has multiple peaks and higher noise levels than the US equivalent. Therefore, we estimate the PA centre frequency by finding the peak of the PA spectrum and centring the band-pass filter around this frequency. From the RF PA data, we estimate the centre frequency to be between 5.13 and 5.61~MHz, so we have used the average of 5.37~MHz for these experiments. 

The PA data are band-pass filtered with frequency cut-offs 2~MHz above and below the centre frequency (bandwidth $\Delta f= 4$~MHz). The PA axial resolution (AR) is given by $0.88 \times c/\Delta f=0.33$~mm~\cite{jeon_review_2019}, which is equal to the US AR, found as half the duration of the strongest reflection off of the PMMA phantom (after fast-time band-pass filtering using the cut-off frequencies listed in Table~\ref{table_params}), divided by the SOS. We estimate the lateral resolution in the vector flow images as $1.2\times \text{F-number}\times\lambda$~\cite{lu_biomedical_1994}, and these values are stated in Table~\ref{table_params} for pixels within the channel ($c=1485$~m/s, $f_{0,\text{PA}}= 5.37$~MHz, $f_{0,\text{US}}= 7.60$~MHz).

\subsection{Refraction-corrected delay-and-sum with multi-angle observations}\label{multi_angle}
\color{black}
\begin{figure*}
    \centering
    \includegraphics{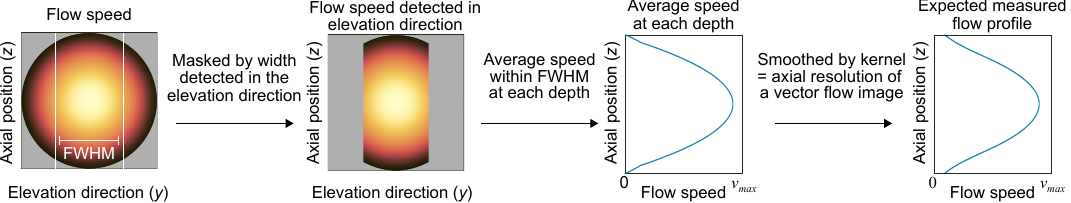}
    \caption{Process to calculate the expected flow speed profile for PAVF and USVF imaging. We estimate the FWHM distance for PA and US in the elevation direction via Field II simulations. In the two left panels, the colourmap indicates the flow speed at the different locations in the channel cross-section, with yellow indicating faster flow speeds than black.}
    \label{expected_flow_speed_calc}
\end{figure*}
 
We create multi-angle observations at each pixel by reconstructing the $I/Q$ data from discrete receive subapertures (Rx) using RC-DAS \cite{renaud_vivo_2018, salles_revealing_2021, shepherd_photoacoustic_2020}. The pixel size is 50~$\mu m$ by 50~$\mu m$. For each Rx, the receive angle $\phi$ at the pixel is chosen for all pixels. The corresponding receive aperture is obtained by tracing the rays back from the pixel to the probe, while accounting for the refraction at each interface, as shown in Fig.~\ref{geo}. The number of elements in the subaperture is given by the user-defined F-number. Through empirical optimisation, we found an F-number of 4 was appropriate, as F-numbers between 2-6 all gave similar results.  


We used five Rx with $\phi$ ranging between $\pm~10^\circ$ for the multi-angle image reconstruction of both PAVF and USVF images. Typically, the $\phi$ range is limited to $\pm~10^\circ$ for a linear array where the pitch is approximately one wavelength~\cite{yiu_least-squares_2016}. In our previous work in single SOS media, we also used five Rx as this was found to provide an adequate number of observations for accurate velocity estimates while maintaining reasonable computation times~\cite{smith_vector-flow_2024}. 

For USVF, we also utilise steered plane waves. The angle of these plane waves is chosen by the operator, and the transmit delays at each element to generate the steered plane waves are calculated using the lens SOS. When a plane wave passes through the lens (and each new interface), the plane wave is refracted, so the angle of the plane wave at a pixel ($\theta$) will differ from the angle of the plane wave that was chosen in the lens ($\theta_l$). Fig.~\ref{geo} shows the difference between $\theta$ and $\theta_l$. In \eqref{matrix_vf_US}, $\theta$, not $\theta_l$, is used to find the velocity at a pixel.

\subsection{Clutter filter}
We clutter filter the US and PA $I/Q$ images to suppress the stationary signal and highlight the frame-to-frame variability caused by the flowing particles. We use a fourth-order Butterworth filter with a cut-off frequency equal to $0.01\times $FR$/2$, or 1\% of the Nyquist frequency. This cut-off was chosen to suppress the DC peak seen in the slow-time spectra (Fig.~\ref{slow_time_spectra}a,c). The PA spatial period is twice that of the US spatial period due to the one-way propagation of PA compared to US. Therefore, when the US FR is twice the PA FR, and the PA and US centre frequencies are the same, these high-pass filter cut-off frequencies correspond to the same flow velocity in the direction of the beam.

\subsection{Estimating the velocity at each pixel}

For each Tx/Rx combination, we obtain an image showing the average phase-shift $\overline{\Delta\Psi}$ over slow-time using lag-one autocorrelation. These multi-angle estimates of $\overline{\Delta \Psi}$ for each Tx/Rx are input into \eqref{matrix_vf_PA} and~\eqref{matrix_vf_US} and solved using the least squares approach to estimate $U_x$ and $U_z$ (hence, $v_x$ and $v_z$) for every pixel. After the velocity components at each pixel are obtained, the vector flow images are smoothed using a 5~pixel~$\times$~5~pixel kernel, a distance which is equal to 0.25~mm.

We generate a mask to select a region of interest inside the channel to compare the velocity estimates between different phantoms and reconstruction approaches. These masks are determined using the PA B-mode images to locate the upper channel wall. Given that the channel diameters are known, the lower wall of the mask is separated from the upper wall by this distance. We shift the mask in the $z$-direction until the flow profile is centred in the channel. The same mask is used for both the PA and US vector flow images processed with the same phantom and image reconstruction approach. An alternative approach if the channel location is unknown would be to use the power Doppler signal to define the mask.

\subsection{Expected flow profile}\label{find_exp_speed}

To evaluate the accuracy of the flow estimates obtained with PAVF and USVF, we model the flow profile we expect to obtain for the two modalities, which is outlined in Fig.~\ref{expected_flow_speed_calc}. We assume a parabolic flow profile with laminar flow since the flow speeds are low and that the centre of the flow channel is along the US imaging plane. We are using a linear transducer, and the elevation width is on the order of the channel diameter, so the 3D flow information is compressed onto a 2D plane with PAVF and USVF~\cite{smith_vector-flow_2024}. Additionally, we assume the plane of the US probe is aligned with the middle of the flow channel. The factors that contribute to the expected flow profile are described below.

\subsubsection{Elevation width}
We used Field II to simulate the receive and transmission beam profiles in the elevation direction for the L11-5v transducer in a homogenous medium \cite{jensen_calculation_1992, jensen_ja_field_nodate}. The FWHM distance detected in the elevation direction at $z = 21$~mm is 0.8~mm for US and 1.1~mm for PA. The US FWHM distance is narrower due to focusing in both transmit and receive. As a result, these two modalities probe different widths of the channel cross-section, so we expect PA and US to estimate different average flow speeds for the same flow phantom.

\subsubsection{Axial resolution of the vector flow image}\label{US_axial_res}
\begin{figure}
    \centering
    \includegraphics{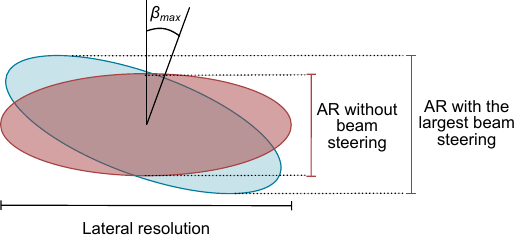} 
    \caption{The resolution within images acquired with beam steering ($\beta\neq0^\circ$, blue oval) is related to the axial and lateral resolution of the unsteered images (red oval). The largest axial resolution (AR) in the images contributing to a vector flow image corresponds to the largest $\beta$ in the image PSF due to beam steering ($\phi$ and $\theta$).}
    \label{vf_axial_res}
\end{figure}

\color{black}
The vector flow images have a finite resolution in the axial direction, so signals originating within an axial distance equal to the vector flow AR are averaged together. We can model the central component of the image point spread function (PSF) in an image with no beam steering as an ellipse, with width and height given by the lateral resolution and AR, respectively. However, with beam steering, the image PSF is rotated by an angle $\beta$, and the AR distance increases with $\beta$, as shown in Fig.~\ref{vf_axial_res}. In PA, the PSF rotation is only due to $\phi$, so $\beta_{max}=10^\circ$. For US, the rotation of the PSF is given by $\beta=(\theta+\phi)/2$~\cite{lecomte_resolution_2008}, so for the most extreme $\phi$ and $\theta$ angles in the channel $\beta_{max} =(19^\circ+10^\circ)/2=14.5^\circ$. Since the vector-flow processing utilises images where the magnitude of $\beta$ ranges from 0$^\circ$ to $\beta_{max}$, we estimate that the AR in the vector flow images is the average of the unsteered AR and the AR with the largest beam steering angle, $\beta_{max}$. With this approach, we have determined the AR in the vector flow images to be 0.36~mm for both PAVF and USVF. The expected flow profile is then smoothed with a moving window equal to the AR in the vector flow images.

\section{Results}

\begin{figure*}
    \centering
    \includegraphics{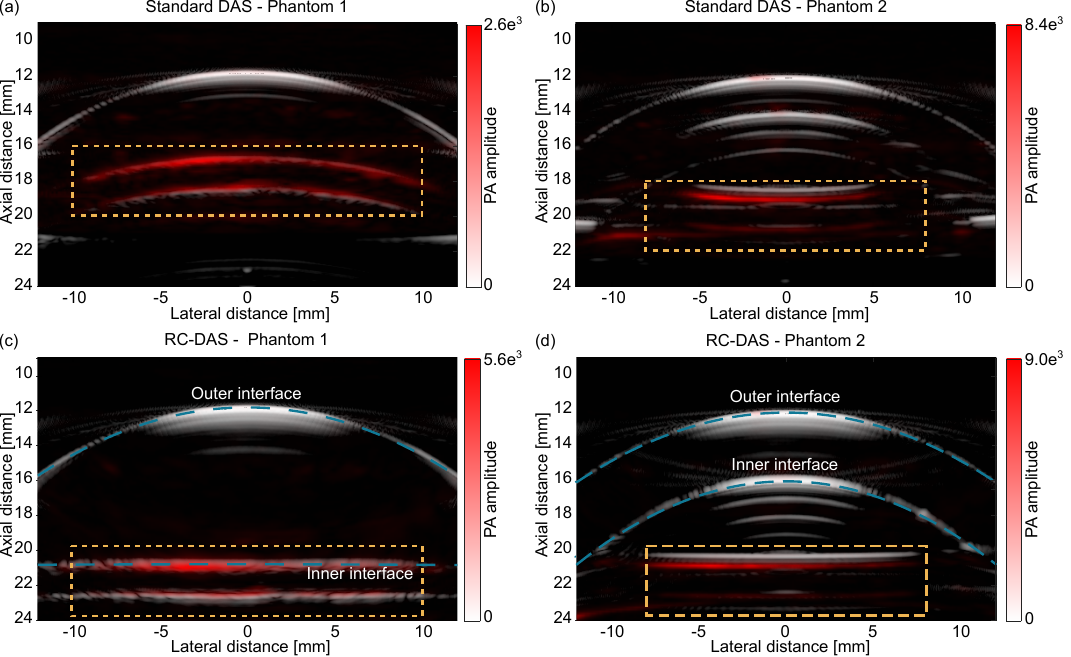}
    \caption{B-mode images for Phantom 1 (a,c) and Phantom 2 (b,d) when using standard DAS (a,b) compared to RC-DAS for image reconstruction (c,d). The grey-scale image shows the US images, while the red overlay shows the PA. The yellow boxes indicate where the vector flow images have been obtained in Fig.~\ref{VF_all}. For images reconstructed with RC-DAS, the interfaces of the PMMA layers are shown by the blue dashed lines labelled ``inner/outer interface''. }
    \label{bmodes}
\end{figure*}

\subsection{B-mode imaging}

\color{black}
Figs.~\ref{bmodes}a-b show the US and PA B-mode images reconstructed with the standard DAS approach assuming a homogeneous medium ($c=1485$~m/s). With standard DAS, the channel in Phantom 2 is curved slightly upwards, while the channel in Phantom 1 is curved downwards. With RC-DAS, straight channels are seen for both phantoms. In the B-mode image of Phantom 2 with standard DAS (Fig.~\ref{bmodes}a), the channel is located 5.1~mm below the outer surface at $x=0$~mm, while with RC-DAS (Fig.~\ref{bmodes}c), the channel is located 9.1~mm beneath the outer surface at $x=0$~mm. The channel depth with RC-DAS is closer to the true distance, which we estimate to be $\sim$9.5~mm. Similarly, in the B-mode images of Phantom 1 reconstructed using standard DAS (Fig.~\ref{bmodes}b), the thickness of the PMMA layer is 2.1~mm at $x=0$~mm, while with RC-DAS (Fig.~\ref{bmodes}d), the thickness of the PMMA is 3.94~mm at $x=0$~mm, which is close to the true thickness of 4~mm measured using vernier callipers.

In the PA B-mode images of Phantom 2, the field-of-view of the channel is wider in the lateral direction when we use RC-DAS. For both phantoms, the PA image pixel amplitudes are higher with RC-DAS, indicating that the image focus has improved. The PA image amplitudes are lower for Phantom 1 than they are in Phantom 2 for the same type of image reconstruction, as there is more PMMA in the former, so there is more acoustic attenuation. In the PA B-mode images, only the wall signals are visible, which is caused by the boundary build-up effect. This effect is common in PA imaging and is caused by the coherence of PA waves at the walls and the limited detection bandwidth~\cite{guo_specklefree_2009, vilov_photoacoustic_2020}. In the B-mode images of Phantom 2 (Figs.~\ref{bmodes}b, d), there are US signals reflected by both the outer and inner tube walls, and the US signal from the inner wall is aligned with the PA signal generated in the carbon suspension at the inner wall.

In the B-mode images of Phantom 2 (Figs.~\ref{bmodes}b, d), there are US signals reflected by both the outer and inner tube walls, and the US signal from the inner wall is aligned with the PA signal generated in the carbon suspension at the inner wall.


\subsection{Extraction of flow signal}

\begin{figure*}
    \centering
    \includegraphics{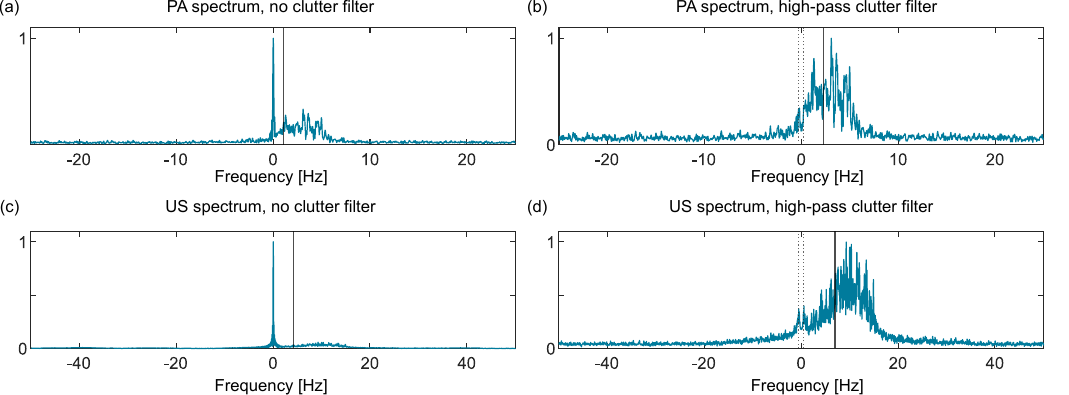}
    \caption{Average slow-time spectra from a 20$\times$20 pixel region in the middle of the channel with PAVF (a,c) and USVF (b,d) for Phantom 1. The receive angle for these spectra is -10$^\circ$, while $\phi_l$ for the US spectra is also -10$^\circ$. The spectra before (a,c) and after (b,d) clutter filtering with the high-pass filter are shown. The black dotted lines show the cut-off frequencies of the clutter filter. The solid black lines show the centre-of-mass frequency found from the mean phase-shift obtained using lag-one autocorrelation since $f_{COM}=\overline{\Delta\Psi}\times \text{FR}/2\pi$.}
    \label{slow_time_spectra}
\end{figure*}

Fig.~\ref{slow_time_spectra} shows the average slow-time spectra for a $20~\times~$20-pixel area in the middle of the channel before and after clutter filtering. Without clutter filtering, both the PA and US spectra are dominated by signal clutter, which presents as signal centred at 0~Hz in Fig.~\ref{slow_time_spectra}a-c. In the PA spectra, there is also a significant signal due to flow, which is centred around $\sim 2.5$~Hz. The signal-to-clutter ratio is significantly higher in the PA spectra than for US, as evidenced in Fig~\ref{slow_time_spectra}a,b. This implies that PA is less reliant on clutter filtering than US to extract the flow signal, which may be beneficial in slow-flow situations where the clutter and flow signals overlap spectrally and are challenging to separate.

\begin{figure*}
    \centering
    \includegraphics{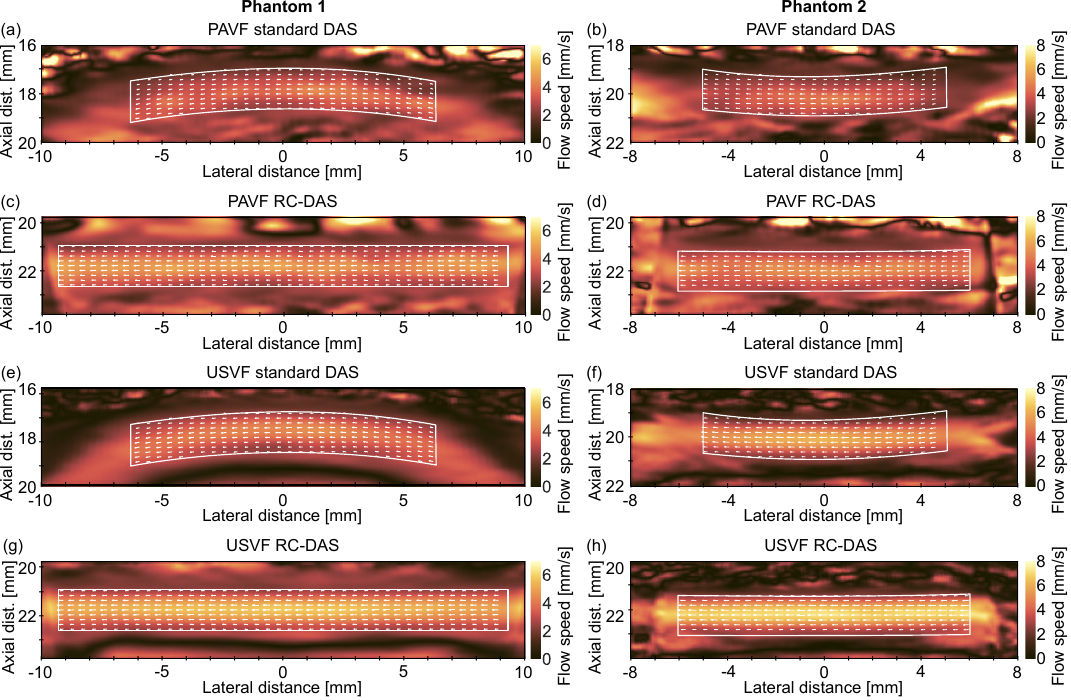}
    \caption{PAVF and USVF images generated with data reconstructed using standard DAS and RC-DAS for the two phantoms. The colour indicates the flow speed, and the arrows show the velocity inside the white box. The mask indicates the area inside the channel where coherent flow estimates are obtained. The average speeds inside the masked areas are shown in Table~\ref{flow_speeds_table}.}
    \label{VF_all}
\end{figure*}

\subsection{Velocity estimates} \label{results_velocity_est}

\color{black}

\begin{figure*}
\centering
\includegraphics{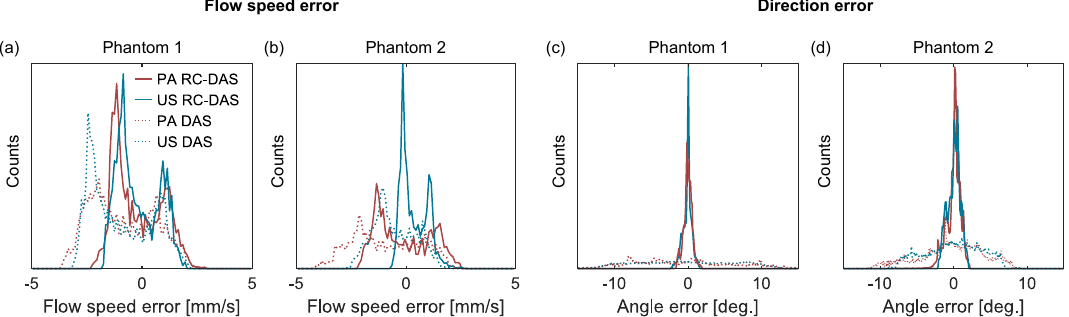}
\caption{Histograms showing the errors in the flow speed estimates with DAS and RC-DAS (a-b) from pixels inside the masked regions in Fig.~\ref{VF_all} compared to expected flow profiles. Histograms showing the errors in direction estimates of vectors inside the masked regions in Fig.~\ref{VF_all} compared to the known angle of the channel found from the B-mode images are shown in (c-d). For Phantom 1, the mode angle estimates with RC-DAS are located at an error of -0.15$^\circ$ and 0.05$^\circ$ for PAVF and USVF, respectively. For Phantom 2, the mode angle estimates are located at an error of 0.14$^\circ$ and 0.58$^\circ$ for PAVF and USVF, respectively. }
\label{speed_dir_error_est}
\end{figure*}

\color{black}

Fig.~\ref{VF_all} shows the PAVF and USVF images for both phantoms with standard DAS and RC-DAS. The colour indicates the flow speed estimated at each pixel, and the arrows show the velocity vector for points inside the channel. The white boxes indicate the masks used to isolate the pixels inside the channel, and the average flow speeds inside the masks are summarised in Table~\ref{flow_speeds_table}. Flow speed estimates from outside of the channel are meaningless since there is no flow present there. This false flow signal is due to noise that remains after clutter filtering or due to the extended PSF, which smears the appearance of flow beyond the channel. We could use power Doppler to mask these signals, however, we have included the entire field of view for transparency.
\color{black}

We quantify the error between the flow speed estimates with PAVF/USVF and the expected flow speeds inside the channel (estimated in Section~\ref{find_exp_speed}) using the mean absolute error (MAE, see Table~\ref{flow_speeds_table}). We chose the MAE metric over the more common root mean square error (RMSE) as MAE is more robust to outlying values since the RMSE squares the error, which gives outliers more influence.

\subsubsection{Flow speed estimates}

When we use standard DAS reconstruction (Fig.~\ref{VF_all}a-b,e-f), the parabolic flow profile expected for laminar flow is only visible near $x=0$~mm where the phantom surface is parallel to the probe and becomes less parabolic with increasing distance from $x=0$~mm. As such, the masked region where coherent flow estimates are located extends further in the lateral direction for both phantoms when we use RC-DAS compared to standard DAS (Fig.~\ref{VF_all}c-d,g-h). 
\color{black}

For both phantoms and modalities, RC-DAS increases the flow speed accuracy, as shown by the decrease in the MAE when RC-DAS is used compared to standard DAS (see $\Delta$MAE column in Table~\ref{flow_speeds_table}). In Phantom 1, the MAE decreases by 0.45~mm/s for PAVF, and by 0.63~mm/s for USVF. For Phantom 2, the MAE decreases by 0.41~mm/s for PAVF, and by 0.43~mm/s for USVF. 

The flow speed errors (difference between USVF/PAVF estimates and the expected flow speed) at each pixel are plotted as histograms in Fig.~\ref{speed_dir_error_est}a-b. For both modalities, the mode flow speed errors (shown as the peak values in Fig.~\ref{speed_dir_error_est}a-b) shift towards 0~mm/s, and the range of the flow speed errors decreases when RC-DAS is used. In Phantom 1 with RC-DAS, the mode flow speed error in Fig.~\ref{speed_dir_error_est}a shifts from -2.00~mm/s to -1.16~mm/s with PAVF, and from -2.42~mm/s to -0.87~mm/s with USVF. In Phantom 2 with RC-DAS, the mode flow speed error in Fig.~\ref{speed_dir_error_est}b shifts from -2.10~mm/s to -1.33~mm/s with PAVF, and from -1.05~mm/s to -0.16~mm/s with USVF.
\color{black}

\begin{figure*}
    \centering
    \includegraphics{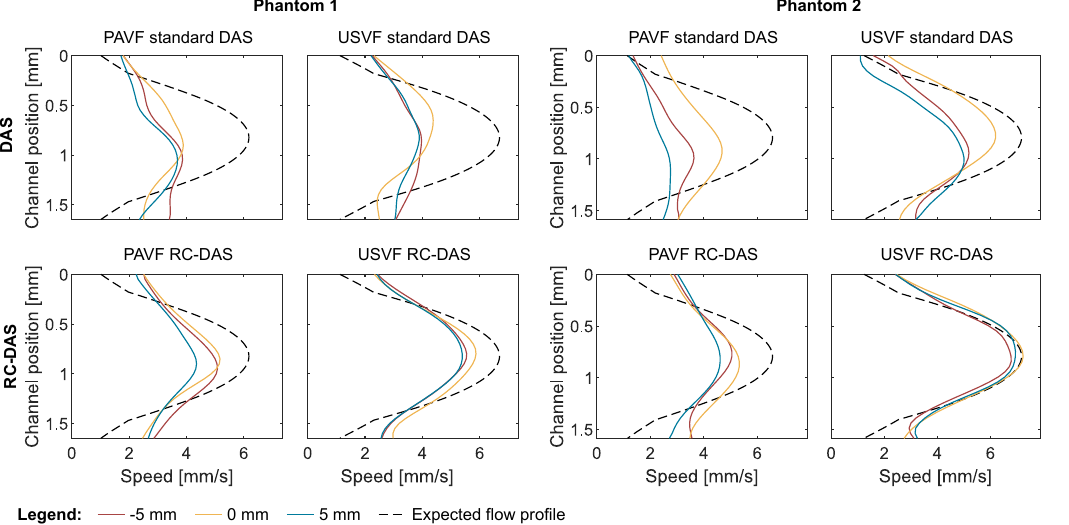}
    \caption{Flow speed profiles at different lateral positions in the masked region shown in Fig.~\ref{VF_all} for the PAVF and USVF results in Phantom 1 (left columns) and Phantom 2 (right columns). The top row shows the profiles with the standard DAS, while the bottom row shows the profiles with RC-DAS reconstruction. The coloured lines show the estimated flow profile at lateral positions: -5~mm (red), 0~mm (yellow) and 5~mm (blue), while the black dashed line indicates the expected profile found in Section~\ref{find_exp_speed}.}
    \label{flow_profile}
\end{figure*}

Fig.~\ref{flow_profile} shows the flow profile with depth at different lateral positions $x=$~[-5~mm, 0~mm, 5~mm]. We compared these profiles to the expected average flow profile (black dashed line). For standard DAS, we obtain the most accurate estimates at $x=0$~mm, while RC-DAS improves the shape and similarity of the flow profiles at all three lateral positions. From both Fig.~\ref{flow_profile} and the vector flow images in Fig.~\ref{VF_all}, we observe that the maximum flow speeds from the centre of the channel increase to become more similar to the expected flow speed estimates with RC-DAS. For USVF with RC-DAS, the maximum flow speeds in Phantom 2 are closer to the expected flow profile than in Phantom 1. The explanation for this difference between the two phantoms is discussed in Section~\ref{US_flow_speed_disc}.

\begin{table*}
\caption{Flow speed estimates and errors}
\centering
\begin{tabular}{ll|l|ll|ll|l}
\hline
\hspace{3.5cm} & True average flow speed (3D) & Expected average flow speed (2D) & DAS & MAE & RC-DAS & MAE & $\Delta$MAE \\ \hline
Phantom 1, PAVF & \multirow{2}{*}{$3.70~\pm~0.28$} & 3.92 & 3.02 & 1.46 & 3.71 & 1.01 & -0.45 \\
Phantom 1, USVF &  & 4.34 & 3.38 & 1.49 & 4.14 & 0.86 & -0.63 \\
Phantom 2, PAVF & \multirow{2}{*}{$3.99~\pm~0.07$} & 4.17 & 3.22 & 1.52 & 4.04 & 1.11 & -0.41 \\
Phantom 2, USVF &  & 4.65 & 4.10 & 0.96 & 5.01 & 0.53 &  -0.43\\ \hline
\multicolumn{8}{c}{}\\
\multicolumn{8}{p{490pt}}{Magnitude of average flow velocities (in units of mm/s) for the two phantoms inside the masked area with both PAVF and USVF imaging. The ``True average flow speed (3D)'' describes the mean speed of the fluid flowing through the channel based on the pump rate and inner diameter of the channel. The ``Expected average flow speed (2D)'' describes the expected average flow speed inside the channel found in Section~\ref{find_exp_speed}. The columns labelled ``DAS'' and ``RC-DAS'' show the average flow speed inside the mask when reconstructed with these approaches. To the right of these values are the mean absolute errors (labelled as ``MAE'') that describe the average absolute error relative to the expected average flow speed at each pixel inside the masked area. The column labelled ``$\Delta$MAE'' shows the change in MAE value when RC-DAS is used compared to standard DAS.}
\end{tabular}
\label{flow_speeds_table}
\end{table*}

\color{black}

\subsubsection{Direction estimates}

\begin{table}
\centering
\caption{Flow direction estimate errors}
\begin{tabular}{@{}llllll@{}}
\toprule
  & &\multicolumn{2}{c}{Phantom 1} & \multicolumn{2}{c}{Phantom 2} \\ 
 & & Average error & IDR & Average error & IDR \\\midrule
PA DAS & & -0.45$^\circ$ & 18.7$^\circ$ & -0.26$^\circ$ & 12.2$^\circ$ \\
US DAS & &-0.27$^\circ$ & 18.0$^\circ$ & -0.03$^\circ$ & 10.5$^\circ$ \\
PA RC-DAS & & -0.01$^\circ$ & 1.39$^\circ$ & -0.02$^\circ$ & 2.38$^\circ$ \\
US RC-DAS & & -0.10$^\circ$ & 1.53$^\circ$ & 0.04$^\circ$ & 2.37$^\circ$ \\ \hline
\multicolumn{6}{c}{}\\
\multicolumn{6}{p{230pt}}{The columns labelled ``Average error'' show the difference between the mean flow angle in the channel with PAVF/USVF and the angle of the channel found from the B-mode images with RC-DAS. The column labelled ``IDR'' describes the interdecile range, found as the difference between the 10th percentile and the 90th percentile. This metric describes the spread of the angle estimates.}
\end{tabular}
\label{angle_tab}
\end{table}

The angle of the flow vectors inside the masked regions are plotted as histograms in Fig.~\ref{speed_dir_error_est}c-d relative to the direction of the channel found from the RC-DAS B-mode images. The average angle errors and interdecile range (IDR) of the angle estimates for each vector flow image are shown in Table~\ref{angle_tab}. The IDR describes the spread of the angle estimates, given as the difference between the 10th and the 90th percentiles.

As shown in Fig.~\ref{VF_all}, when the images are reconstructed with standard DAS, the flow vectors point in a curved path that follows the curvature of the channels seen in the B-mode images (Fig.~\ref{bmodes}a-b). When we use RC-DAS, the flow vectors inside the channel become more unidirectional, which is shown by a decrease in the flow angle IDR. From the IDR values in Table~\ref{angle_tab}, for Phantom 1 with RC-DAS, the IDR reduces by 17.3$^\circ$ for PAVF, and by 16.5$^\circ$ for USVF. Similarly, for Phantom 2 with RC-DAS, the IDR reduces by 9.82$^\circ$ for PAVF, and by 8.13$^\circ$ for USVF. With RC-DAS, the IDR is larger for Phantom 2, as the tubing is not perfectly straight, so there is a wider range of direction estimates present. 

\section{Discussion}

\subsection{B-mode imaging}

As shown in Fig.~\ref{bmodes}, RC-DAS image reconstruction more accurately maps the time-series pressure data to the 2D imaging grid compared to the standard DAS image reconstruction. This allows the correct geometry to be obtained in the PA/US images and increases the contrast of pixels in the B-mode images. The lateral positions furthest from $x=0$ correspond to where the outer PMMA surface is the steepest, resulting in the most extreme refraction and phase aberration. As a result, the RC-DAS image reconstruction increases the length of the channel that can be seen with PA by accurately reconstructing these highly refracted signals. This is why the masked area is longer in the lateral direction when RC-DAS reconstruction is used for the same phantom. 

\subsection{Flow speeds estimates}

\subsubsection{Influence of RC-DAS on the axial flow speed profile}\label{RC-DAs_speed_exp}
With standard DAS, the $I/Q$ images used to obtain the vector flow images are ``blurred'' due to incorrect time delays being used in the reconstruction as a result of assuming a homogenous SOS model. Therefore, the flow profiles obtained with standard DAS are more uniform with depth inside the channel. This is because slower flow speeds originating near the channel walls are mapped to the middle of the channel, lowering the $\overline{\Delta\Psi}$ estimates and flow speeds detected in the middle of the channel. This can be seen in Fig.~\ref{flow_profile}, where the parabolic flow profile is less evident compared to when we use RC-DAS.

\subsubsection{Underestimation near the channel walls}
Near the top and bottom edges of the channel, the PAVF and USVF flow speed estimates are overestimated compared to the expected flow profile (Fig.~\ref{flow_profile}). Our expected flow speed profile does not taper to 0~mm/s at the walls, as we have a finite AR, meaning flow speeds are averaged in the axial direction in the vector flow images~\cite{yiu_vector_2014}. As a result, pixels near the wall will detect faster flow that originates from closer to the middle of the channel. However, due to propagation through the layered media, we hypothesise that the PSF is in fact larger than the AR calculated in Section~\ref{vf_axial_res}, i.e. the window size used to smooth the expected flow profile is not large enough to accurately model the flow profiles that PAVF and USVF would detect.

\color{black}


The histograms showing the flow speed errors in Fig.~\ref{speed_dir_error_est}~a-b are bimodal, with one peak located in the negative flow speed error region (slower than expected) and one peak in the positive flow speed error region (faster than expected). We can deduce from the flow profiles in Fig.~\ref{flow_profile} that these negative-error peaks in Fig.~\ref{speed_dir_error_est}~a-b are due to the underestimation of the flow speeds near the middle of the channel. As described in Section~\ref{RC-DAs_speed_exp}, RC-DAS reduces these errors and shifts the peaks in Fig.~\ref{speed_dir_error_est}a-b towards an error of 0~mm/s. However, even with RC-DAS, there is still some residual flow-speed error, which is discussed in Sections~\ref{PA_flow_speed_disc} and \ref{US_flow_speed_disc} for PA and US, respectively. We hypothesise that the positive-error peak is due to the finite AR described above, which causes slow-moving pixels near the walls to be averaged with the faster flows deeper in the channel. Unlike the negative-error peak, the positive-error peak is unchanged with RC-DAS, as the source of the overestimation near the channel walls remains.

\subsubsection{PAVF flow estimates}\label{PA_flow_speed_disc}

Even with RC-DAS, the peak PA flow speeds from the middle of the channel are slower than those found in the expected flow speed profile, and the MAE values are higher for PAVF than USVF (Table~\ref{flow_speeds_table}). The PA signal is proportional to the optical fluence, which varies due to optical scattering and absorption by the carbon suspension within the tubing cross-section. As described in Section~\ref{find_exp_speed}, the 3D flow information is compressed onto a 2D imaging plane. Therefore, the flow near the out-of-plane edges of the channel will contribute more to the detected flow speeds at a pixel since the fluence is higher there, which explains why the flow near the middle of the channel is slower than expected. To address this issue and improve the expected flow profile, one could weight the flow speeds estimated with PAVF by the fluence in the elevation direction, as shown in our previous work \cite{smith_vector-flow_2024}. Additionally, PAVF uses half as many frames as USVF, so the slow-time spectra are noisier (Fig.~\ref{slow_time_spectra}). This biases the $\overline{\Delta\Psi}$ estimates towards zero, lowering the velocity estimates found using PAVF. 

\subsubsection{USVF flow estimates} \label{US_flow_speed_disc}
Even with RC-DAS, the velocity estimates from the middle of the channel in Phantom 1 are slower than the expected flow profile, whereas the RC-DAS estimates in Phantom 2 are in better agreement with the expected flow profile. One reason for this is that the Field II simulation used to estimate the FWHM of the elevation direction at the depth of the channel with PA and US did not consider the presence of the PMMA. The elevation focus depth of the L11-5v probe is 18~mm, however, due to refraction with the PMMA layer, the true elevation focus depth would be closer to the probe. As a result, the actual FWHM size detected in the elevation direction at the channel depth would be larger than the value used to calculate the expected average flow speed. This difference in the elevation FWHM is more significant in Phantom 1 since there is more PMMA present, which explains why the peak flow speeds estimated with USVF are slower than expected for Phantom 1, while Phantom 2's results are closer to the expected profile.

Another reason why the speed estimates in Phantom 1 are slower than the expected profile in the middle of the channel is that $f_0$ used in the vector flow calculations is not representative of the signal from within the channel due to attenuation by the PMMA, resulting in the fast-time spectra from the channel becoming skewed with depth, decreasing $f_0$. As such, using the centre frequency of the detected signal in \eqref{pa_dopp_1} and \eqref{US_dopp_1} may lead to the flow speeds being underestimated. This effect is pronounced in Phantom 1 since there is more PMMA present. To overcome this, we could estimate $f_0$ at each pixel by estimating the spatial period in the images~\cite{loupas_axial_1995}. Finally, the results for Phantom 1 could be slower than expected if the tubing was not perfectly aligned along the US imaging plane. While, we tried to ensure good alignment using real-time B-mode imaging to enhance the detected amplitude, if the channel was slightly misaligned, the centre of the flow channel would be shifted into the out-of-plane direction, so these VFI techniques would be less sensitive to the fastest flow in the middle of the channel.


\subsection{Phantom limitations} 
A limitation of our phantom experiments is that the flow is always in the lateral direction, however, we have previously demonstrated PAVF and USVF in an angled soft-tissue flow phantom~\cite{smith_vector-flow_2024}. The main challenge with angled flow is that aliasing occurs at slower speeds as the flow component in the direction of the beam is increased with flow angle. However, there are dealiasing techniques to overcome this issue~\cite{ekroll_extended_2016,poree_dealiasing_2021, nahas_gpu-based_2023}.

\subsection{Potential Applications}
The techniques described in this manuscript could be applied to quantify blood flow within and beneath bone tissue to provide insights into bone health, disease progression, and treatment efficacy~\cite{dyke_noninvasive_2010}. Additionally, these techniques could be extended for transcranial imaging~\cite{mozaffarzadeh_refraction-corrected_2022}. However, several challenging factors need to be addressed for successful in vivo translation. For in vivo VFI in bone, both the acoustic attenuation~\cite{szabo_appendix_2014, ono_comprehensive_2020} and the optical attenuation will be greater than in our PMMA phantoms. This will result in decreased signal-to-noise ratio (SNR) from blood for both modalities, which may reduce the imaging depth and decrease the magnitude of $\overline{\Delta\Psi}$ estimates, reducing the magnitude of the velocity estimates.


Beyond biomedical applications, these VFI techniques could also be applied to NDT for fluid flow quantification. In engineered systems, multilayered media with high SOS contrast are more common, and the ability to resolve the flow profile within such structures could be valuable for monitoring fluid transport~\cite{tezuka_ultrasonic_2008, murakawa_dealiasing_2015, a_nour_review_2020, krishna_ultrasound_2022}.

\subsection{Choice of imaging modality}
PA and US imaging modalities may each offer distinct advantages for VFI depending on the specific imaging context. In general, US imaging is well-suited for measuring faster flow speeds in the axial direction ($>$~1~cm/s) due to the high frame rates on the order of $\sim$10~kHz achievable with ultrafast US~\cite{montaldo_coherent_2009, bercoff_ultrafast_2011}. In contrast, PA imaging is limited by laser repetition rates -- typically in the tens to hundreds of Hz -- due to safety regulations~\cite{laser_standard} and compromises between pulse energy and repetition rate. Therefore, there is a trade-off between pulse energy, SNR, and imaging depth with FR~\cite{zhang_photoacoustic_2024} for PA imaging, which ultimately limits the maximum measurable flow rates. The pressure signals generated by US transducers are higher (MPa) than laser-generated PA signals (kPa)~\cite{beard_biomedical_2011}; therefore, US is generally more suited for imaging deeper targets, or in situations with poor optical penetration. In US imaging of blood, the signal from stationary and slowly-moving tissues -- referred to as the clutter signal -- is significantly stronger than the signals from the blood particles, so clutter filtering is required. However, it is difficult to isolate the flow signal when the flow rates are slow, as this signal will spectrally overlap with the clutter signal~\cite{jansson_estimation_1999,cosgrove_imaging_2010, heimdal_ultrasound_1997, seddiki_advancements_2025}. In contrast, the specificity of PA imaging may be beneficial for slow-flow, as the clutter signal can be reduced by selecting an appropriate wavelength to excite a target chromophore that reduces the signal clutter from other components in the sample. In media that are highly acoustically attenuating, such as in bone~\cite{gonzalez_photoacoustic_2023}, PA is advantageous as only the return signal travels as an acoustic wave, so the optical losses in the media are not as significant as the acoustic losses. Finally, PA imaging is promising for quantifying hypoechoic fluid that is not strongly scattered by US but is a strong optical absorber.

\subsection{Outlook for dual-modality VFI systems}
In this work, we have demonstrated that both PA and US can independently perform VFI in layered media. We hypothesise that a dual-modality VFI system would provide redundancy of flow information and may extend the range of flow speeds measurable due to the inherent strengths of each modality, which are outlined in the introduction. Beyond purely VFI applications, dual-modality PA and US imaging can provide complementary information. For example, PA imaging in layered media with RC-DAS requires that the SOS model is known, which requires US imaging. Additionally, PA imaging can be used to obtain spectroscopic information, such as blood oxygen saturation, which US imaging is not sensitive to.

\section{Conclusion}
We present a method for quantifying particle flow with PAVF and USVF in layered media with a high SOS contrast using refraction-corrected image reconstruction combined with a multi-angle vector flow algorithm. This technique is validated by imaging in multi-layered models, and the results are compared to standard DAS. By correcting for refraction, the geometry in the B-mode images is more accurate and the field of view where flow estimates can be obtained increases. Furthermore, the velocity and direction estimates more closely match the expected flow profile with RC-DAS. Our USVF estimates are more accurate for the phantom with a thinner 4~mm PMMA layer (Phantom 2) compared to the thicker 10~mm PMMA phantom (Phantom 1) due to reduced attenuation and aberration, more accurate estimates of the elevation width, or better alignment between the channel and the US imaging plane. This work emphasises that accounting for refraction when estimating flow within layered models is critical to obtaining accurate estimates of both flow speed and direction. While PAVF and USVF can both quantify flow, the most appropriate modality for a given situation will depend on the properties of the media and the flow dynamics.

\section*{Acknowledgment}
The authors would like to thank the Technical Services workshop at the University of Auckland for fabricating the phantoms and acknowledge the Centre for eResearch at the University of Auckland for their help in facilitating this research. Caitlin Smith acknowledges Dodd-Walls Centre for Photonic and Quantum Technologies, New Zealand for her PhD scholarship funding. 

\bibliographystyle{IEEEtran} 
\bibliography{bib2}

\vspace{-1.5cm}

\begin{IEEEbiography}[{\includegraphics[width=1in,height=1.25in,clip,keepaspectratio]{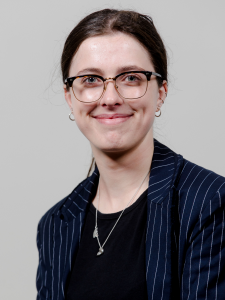}}]{Caitlin Smith} is a PhD student at the University of Auckland. She received her bachelor’s degree in physics and chemistry in 2020 and an honours degree in physics in 2021, both from the University of Auckland. Her research focuses on using ultrasound and photoacoustic imaging to quantify the haemodynamics in bone. 
\end{IEEEbiography}

\vspace{-1.5cm}

\begin{IEEEbiography}[{\includegraphics[width=1in,height=1.25in,clip,keepaspectratio]{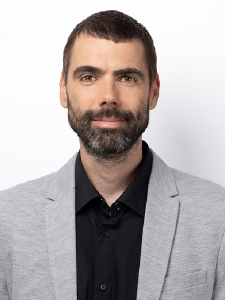}}]{Guillaume Renaud} received the engineering degree from ICAM Nantes, France, in 2004, the Master degree in acoustical physics from the University of Le Mans, France, in 2005, and the Ph.D. degree in biomedical physics from the University of Tours, France, in 2008. After a Post-Doctoral Fellowship at Erasmus MC, Rotterdam, The Netherlands, he was hired as a tenured scientist at CNRS in France, based within the Laboratory of Biomedical Imaging, Paris. He currently works as an assistant professor at Delft University of Technology, The Netherlands. His current research interests include nonlinear acoustics, ultrasound imaging and quantification of blood flow inside and behind bone tissue, and the assessment of bone tissue structural quality with ultrasound imaging.
\end{IEEEbiography}

\vspace{-1.5cm}

\begin{IEEEbiography}[{\includegraphics[width=1in,height=1.25in,clip,keepaspectratio]{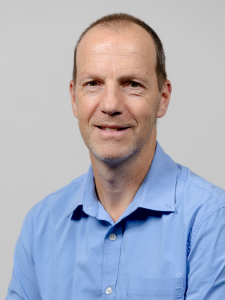}}]{Kasper van Wijk} received his PhD in Geophysics from Colorado School of Mines, and was an associate professor at Boise State University until 2012. In 2012, he joined the physics department of the University of Auckland.
\end{IEEEbiography}
\vspace{-1.5cm}
\begin{IEEEbiography}[{\includegraphics[width=1in,height=1.25in,clip,keepaspectratio]{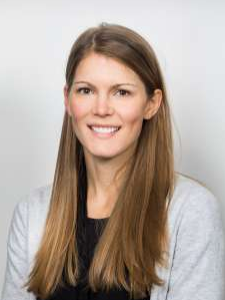}}]{Jami Shepherd} is a Senior Lecturer in the Department of Physics at the University of Auckland.  She is interested in developing new technical capabilities for biomedical photoacoustic, ultrasonic, and laser-ultrasonic imaging, including new capabilities for quantifying bone haemodynamics with ultrasound and photoacoustic imaging.
\end{IEEEbiography}

\end{document}